\documentclass[prd,showpacs,preprintnumbers,amsmath,amssymb,twocolumn]{revtex4-2}
\usepackage{graphicx} 
\usepackage{amsthm,amsmath,amssymb}
\usepackage{mathrsfs}
\usepackage{color}
\usepackage{appendix}
\usepackage{hyperref}
\date{\today}
\begin{document}
\title{Impact of Sub-MeV Dark Matter on the Cooling of Pulsating White Dwarfs}

\author{Bo Zhang}
\affiliation{Department of Physics, Anhui Normal University, Wuhu, Anhui 241002, China} 
\author{Lei Feng}
\email{fenglei@pmo.ac.cn}
\affiliation{Key Laboratory of Dark Matter and Space Astronomy, Purple Mountain Observatory, Chinese Academy of Sciences, Nanjing 210033, China} 
\author{Cui-Bai Luo}
\email{cuibailuo@ahnu.edu.cn}
\affiliation{Department of Physics, Anhui Normal University, Wuhu, Anhui 241002, China}

\begin{abstract}
    In our galaxy, white dwarfs inevitably undergo scattering and capture processes with the interstellar diffuse dark matter. The captured dark matter forms a dark halo that eventually evaporates or annihilates. Theoretical pulsation modes and observations of pulsating white dwarfs provide predictions about their evolution. This motivates us to study the impact of sub-MeV interstellar dark matter on the cooling processes of white dwarfs. In this work, we consider the collisions between dark matter and relativistic degenerate electrons inside white dwarfs, numerically calculating the energy input and output results from scattering, capture, evaporation, and annihilation processes. Based on observational data from G117-B15, we conclude that the maximum cooling luminosity of the interstellar sub-MeV dark matter is approximately $\sim 10^{22}$ erg/s, which is insufficient to provide an effective cooling mechanism for white dwarfs. Finally, if future observations detect a pulsating white dwarf in the Galactic center, the potential sensitivity of this scenario could extend to the region $10^{-3}$ MeV $<m_\chi<10$ MeV and $ 6.02\times 10^{-38} \text{cm}^2 >\sigma_0\geq 1.5\times 10^{-40} \text{cm}^2$.
\end{abstract}
\maketitle

\section{Introduction}

Dark matter represents a fundamental problem in modern physics. Celestial objects, characterized by extreme conditions and abundant sources of dark matter that are unattainable in ground-based experiments, serve as effective probes for dark matter detection. If dark matter interacts with standard model particles, celestial objects can capture diffuse dark matter from their surroundings, leading to the accumulation of dark matter halos. These halos may eventually undergo processes such as evaporation and annihilation. During the evolution of celestial objects, the rates of capture, evaporation, and annihilation can reach equilibrium, stabilizing the number of \textbf{captured dark matter particles} \cite{gaisser1986limits,griest1987cosmic}. Captured dark matter can increase the mass of celestial objects, potentially inducing the collapse of neutron stars into black holes\cite{bell2013realistic,goldman1989weakly,bramante2014bounds,kouvaris2011constraining,kouvaris2011excluding,mcdermott2012constraints,guver2014capture}, and it can also modify energy transport mechanisms in the Sun \cite{gould1990thermal,gould1990cosmion,vincent2014thermal,geytenbeek2017effect}. Furthermore, the annihilation of dark matter accumulated within celestial objects can provide energy to sustain their structure \cite{iocco2008dark,hirano2011evolution,john2024dark}, and it may also produce detectable signals, including neutrinos \cite{choi2015search,bell2021searching,adrian2016search,adrian2016limits,aartsen2017search,nguyen2023bounds}, cosmic rays\cite{linden2024indirect,herrera2024probing,feng2014ams} , and gamma rays \cite{batell2010solar,schuster2010terrestrial,bell2011enhanced,feng2016detecting,Leane:2021tjj,Acevedo:2023xnu,Leane:2021ihh,Leane:2017vag,Leane:2024bvh}. 

White dwarfs, known for their dense structure, extremely long lifespans, and relatively low luminosity, have been extensively studied in the context of dark matter capture \cite{isern1992axion,winget1983pulsating,bose2023impact,bell2021improved,niu2018probing,niu2024possible}. Pulsating white dwarfs, a special subclass, exhibit periodic variations in luminosity over minute timescales. The pulsation pattern offers a valuable opportunity for precise asteroseismic investigations of their internal structures \cite{winget2008pulsating,fontaine2008pulsating,althaus2010evolutionary,calcaferro2017pulsating}, enabling accurate determinations of their cooling luminosity \cite{isern1992axion,corsico2001potential,niu2018probing,niu2024possible}. For GeV dark matter, annihilation processes can inject energy into white dwarfs \cite{niu2018probing,bell2021improved}, whereas the evaporation of sub-MeV dark matter can lead to additional cooling \cite{niu2024possible}. Both phenomena provide a means to constrain the interactions between dark matter and standard-model particles using the cooling luminosity of white dwarfs\cite{isern1992axion,corsico2001potential,niu2018probing,niu2024possible}. 

In this study, we investigate the collisions between sub-MeV dark matter and relativistic degenerate electrons inside white dwarfs, focusing on energy transfer during capture, evaporation, and annihilation processes. We identify that single collisions between dark matter and electrons, where the dark matter subsequently escapes the white dwarf, have a substantial impact on energy transfer dynamics. In such cases, dark matter remains in an unbound state both before and after the collision, which we define as the "scattering process"(To avoid confusion with the ''scattering process'' between dark matter and electrons, this paper exclusively uses the term ''collision'' to describe the interaction between dark matter and electrons.).
Numerical calculations reveal that sub-MeV dark matter can extract energy from white dwarfs through scattering, capture, evaporation, and annihilation processes, potentially serving as an additional cooling mechanism. However, this mechanism does not satisfy the observational constraints on excess cooling luminosity imposed by data from the pulsating white dwarf G117-B15A. By extending our analysis to regions with higher dark matter densities, such as the Galactic center, we explore the potential for white dwarfs to constrain interactions between dark matter and standard model particles. 

The structure of this paper is as follows: Section \ref{section:2 WD} introduces the observational data of pulsating white dwarfs and discusses their structure and evolutionary characteristics. Section \ref{section:3 S-CEA process} focuses on the individual processes of dark matter scattering, capture, evaporation, annihilation, and their equilibrium. In Section \ref{section:4 enegry flux}, we analyze the net energy flux associated with dark matter interactions in stellar systems, specifically addressing the contributions from scattering and capture, evaporation, annihilation processes. Section \ref{section:5 result} presents and interprets our results, while Section \ref{section:6 conclusion} provides a concise summary of the study. 

\section{White Dwarfs}
\label{section:2 WD}

\subsection{Pulsating white dwarf theory and observation}

The period change rate of the pulsation mode in pulsating white dwarfs is given by \cite{winget1983pulsating}:
\begin{equation}
    \frac{\dot{P}}{P} \approx -\frac{1}{2}\frac{\dot{T_\text{c}}}{T_\text{c}}+\frac{\dot{R_*}}{R_*},
\end{equation}
where $P$ represents the pulsation period,  $T_\text{c}$ is the core temperature of the white dwarf, and $R_*$denotes its radius.  This period variation rate have been measured through long-term time-series photometric observation, such as G117-B15A\cite{romero2012toward,kepler2020pulsating,bailer2021estimating}, R548\cite{mukadam2013measuring}, L19-2\cite{pajdosz1995non}. However, for several sources the observed rates exceed those predicted by evolutionary models. Such discrepancies may be attributed to extra cooling mechanisms that dissipate thermal energy. Quantitatively, this relationship is expressed as \cite{isern1992axion,corsico2001potential}: 
\begin{equation}
\frac{\dot{P}_{\text{obs}}}{\dot{P}_{\text{The}}} =\frac{L_*+L_{\text{ext}}}{L_*} ,
\end{equation}
where $\dot{P}_{\text{obs}}$ and $\dot{P}_{\text{the}}$ are the observed and theoretically predicted period change rates, respectively, $L_*$ is the luminosity of the white dwarf derived from asteroseismology, and $L_{\text{ext}}$ accounts for the radiative luminosity due to extra cooling mechanisms. Potential explanations include dark matter candidates such as axions \cite{isern1992axion,corsico2001potential}. Furthermore, these results can constrain the interactions between dark matter and stellar matter, expressed as: 
\begin{equation}
    L_{\chi} \leq \left(\frac{\dot{P}_{\text{obs}}}{\dot{P}_{\text{The}}}-1 \right)L_* < 4.75L_{\odot}.
\end{equation}
Table \ref{DAV_data_table} provides a summary of the observational data for the pulsating white dwarf G117-B15A \cite{romero2012toward,kepler2020pulsating,bailer2021estimating}. For the following analysis, we will use the observational data of G117-B15A as a benchmark for numerical calculations. 

\begin{table}
    \centering
    \begin{tabular}{ccc}
        \hline
            ID &  $d\dot{P}_{\text{obs}}/P_{\text{obs}}$ &$d\dot{P}_{\text{the}}/P_{\text{the}}$  \\
        \hline
             G117-B15A & $(5.12\pm 0.82)\times 10^{-15}$ &$1.25\times 10^{-15}$  \\
        \hline
            $M/M_\odot$ &$\text{log}(L/L_\odot)$ &$\log(R/R_\odot)$ \\
        \hline
         $0.593\pm 0.007$&$-2.497\pm 0.030$&$-1.882\pm 0.029$\\
        \hline
        $P_{\text{obs}}$&$P_{\text{the}}$&Distance \\
        \hline
        215.20&210.215&57.37\\
        \hline
    \end{tabular}
    \caption{Data of G117-B15A\cite{romero2012toward,kepler2020pulsating,bailer2021estimating}}
    \label{DAV_data_table}
\end{table}

\subsection{Structure and evolution of white dwarfs}
\label{section_WDstructure}
White dwarfs are composed of a dense C-O core enveloped by a thin outer shell, which accounts for no more than $1\%$ of the total mass \cite{fontaine2001potential}. In this study, we focus on the energy loss in white dwarfs caused by elastic collisions with dark matter. This process primarily involves the dense core, which constitutes the majority of the white dwarf's mass and exhibits densities ranging from $10^6-10^{10} $ $\text{g}/\text{cm}^3$ .In this extreme state, highly degenerate electrons generate sufficient pressure to counterbalance gravitational forces and prevent stellar collapse.  Owing to the exceptionally high thermal conductivity of the electric degenerate state\cite{solinger1970electrical}, we assume a uniform core temperature $T_*$ in the subsequent calculations. While interactions between dark matter and C-O ions in the core could also influence the luminosity and evolution of white dwarfs \cite{bell2021improved}, these effects lie beyond the scope of this study and will not be discussed in detail. 

\begin{figure}
    \centering
    \includegraphics[width=1.0\linewidth]{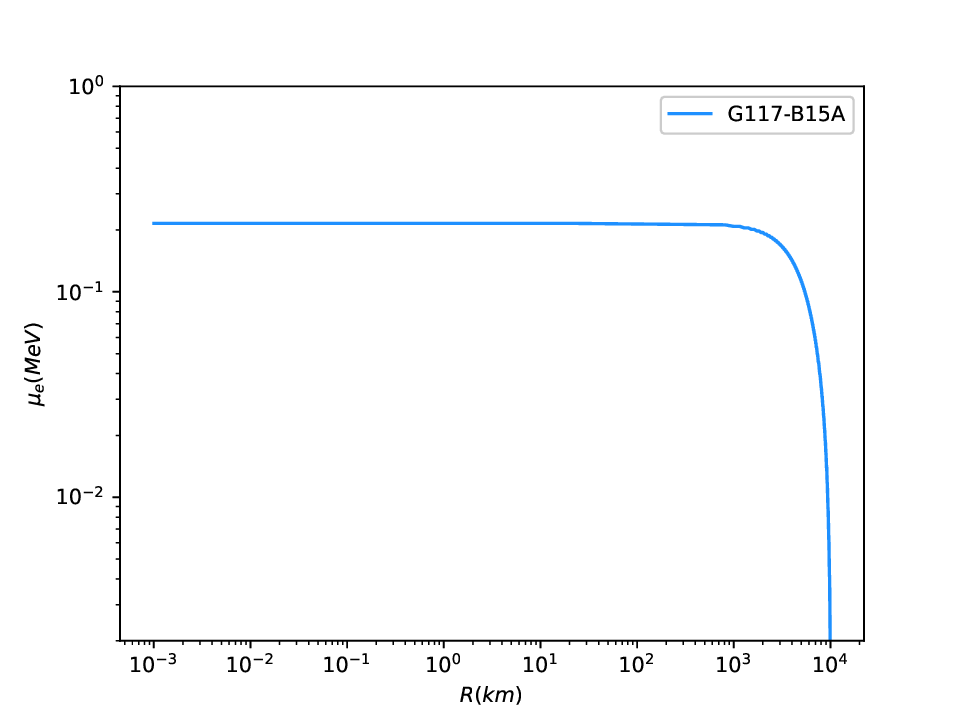}
    \caption{ Electron chemical potential radial profiles for G117-B15A.}
    \label{fig_miu_r}
\end{figure}

The distribution function of electrons in the core of a white dwarf can be described using the Fermi-Dirac distribution:
\begin{equation}
    f_{\text{FD}}(E_\text{e},r)=\left[\exp(\frac{E_\text{e}-\mu(r)}{T_*}) +1\right]^{-1},
\end{equation}
where $E_\text{e}$ denotes the energy of the electrons, $T_*$ represents the core temperature, and $\mu(r)$ is the chemical potential, which varies with the radial distance $r$ within the white dwarf. To obtain the chemical potential, it is necessary to solve the stellar structure equations.  Following the method outlined by Bell \textit{et al.} \cite{bell2021improved}, we employ the Wigner-Seitz cell approximation combined with the assumption of point-like nuclei\cite{salpeter1961energy}, and  the extended Feynman-Metropolis-Teller (FMT) equation of state \cite{feynman1949equations} with incorporating weak interactions and relativistic effects \cite{rotondo2011relativistic1,rotondo2011relativistic2} in the limit of zero-temperature. The structure of white Dwarf can be obtained, assuming a non-rotating spherically symmetric star, by solving the Tolman-Oppenheimer-Volkof (TOV) equations\cite{tolman1939static,oppenheimer1939massive} coupled to the FMT equation of state. In our calculations, the atomic mass number is approximated as the average value of carbon and oxygen, $A=14$. Fig. \ref{fig_miu_r} illustrates the variation of the chemical potential inside the white dwarf, derived from the observational data of G117-B15A. At the center of a white dwarf with a mass of $0.53M_\odot$, the electron chemical potential reaches $0.14$ MeV, indicating that the electrons are relativistic. Consequently, the interactions between dark matter and electrons in the white dwarf core must be analyzed within a relativistic framework. 

The evolutionary process of white dwarfs, from their formation in the explosions of stars with initial masses less than $8-10M_\odot$ to their eventual cooling and dimming, shows a strong correlation between luminosity, core temperature, and age \cite{renedo2010new,salaris2010large}. Utilizing this characteristic, we estimate the core temperature ($1.2 \times 10^7$ K) and age ($1$ Gyr) of G117-B15A. The value of the core temperature is consistent with that used in reference \cite{isern1992axion,corsico2001potential}. 

\section{Scattering, Capture, Evaporate and Annihilation Process}
\label{section:3 S-CEA process}

The capture process increases the dark matter particle population \textbf{in the white dwarf's halo}, while the evaporation and annihilation processes reduce it. Over the course of the white dwarf’s evolution, these competing processes reach an equilibrium, stabilizing the \textbf{captured dark matter particles} density. Notably, we derive general formulas that describe the scatter, capture, evaporation, and annihilation processes, extending their application to a special relativistic framework in section \ref{section:general model}. Based on these formulations, we analyze each process individually in subsequent sections.

\subsection{General model}
\label{section:general model}

Many studies have explored the processes of dark matter capture, evaporation, and annihilation\cite{bell2021improved,garani2017dark,busoni2017evaporation}. These processes can be described by the following general formula: 
\begin{widetext}
\begin{equation}
            \text{Rate} = \int_0^{R_*} 4\pi r^2 dr \int dwdu \left[ n_1(w,r) n_2(u,r) \int dv \frac{d\sigma}{dv} \left|\vec{w}-\vec{u}\right| \Theta(\mathscr{C}(v)) \right],
\end{equation}
\end{widetext}
where $r$ denotes the radial distance within the white dwarf, while $n_1(w,r)$ and $n_2(u,r)$ represent the number density distributions of particles $1$ and $2$, respectively. $w$ and $u$ are the velocities of the particles, $d\sigma/dv$ is the differential cross-section for their collisions, and $v$ is the post-collision velocity of one particle, dark matter normally in our discussion. The step function $\Theta$, combined with $\mathscr{C}(v)$, defines the conditions for scattering events that lead to scattering, capture or evaporation, thereby setting the integration limits. This formula is valid in the non-relativistic regime but becomes inadequate for relativistic particles. For instance, the relative particle flux $F$ is not Lorentz invariant,
\begin{equation}
    F=n_1(w,r)n_2(u,r)|\vec{w}-\vec{u}|.
\end{equation}
Since electrons in white dwarfs are relativistic, as discussed in Section \ref{section_WDstructure}, a relativistic extension is required. To address this, the relative particle flux is replaced by the Møller flux\cite{cannoni2017lorentz}:
\begin{equation}
    F_{\text{møller}} = n_1(w,r) n_2(u,r) \frac{\sqrt{(p_1\cdot p_2)^2-m_1^2 m_2^2}}{E_1 E_2},
\end{equation}
where $m_\text{i},p_\text{i}$ and $E_\text{i}$ are the mass, 4-momentum, and energy of  particle $i$, respectively. In the subsequent sections, we incorporate the relativistic Møller flux into the general formula and combine it with the particle distributions, differential cross-sections, and specific collision conditions to derive the rates for scattering, capture, evaporation, and annihilation processes. 

\subsection{Scattering and capture processes}

When dark matter interacts with electrons inside a white dwarf, it forms a gravitational bound state (capture) if its outgoing velocity is below the escape velocity; otherwise, it forms an unbound state (scattering). Together, these processes equal the total collision rate of dark matter with the white dwarf.  We introduce the scattering rate first, since the capture rate is derived from the scattering rate by modifying the integration range for the outgoing velocity. 

We separately analyze the primary components involved in the scattering process: the degenerate electrons in the white dwarf core and the diffuse interstellar dark matter surrounding it. The distribution of degenerate electrons is discussed in Section \ref{section_WDstructure}, where the electron density distribution is described in terms of the Mandelstam variable $s$ \cite{bell2020improved_neu}:
\begin{equation}
     n_\text{e}(u_\text{e},r)du_\text{e} =  \frac{E_\text{e}dE_\text{e} ds}{4\pi^2 \sqrt{E_\chi^2-m_\chi^2}}f_{\text{FD}}(E_\text{e},r).
\end{equation}
Here, $E_\text{e}$ denotes the relativistic energy of electrons, while $m_\chi$ and $E_\chi$ are the mass and relativistic energy of the dark matter particle, respectively.

From a broader perspective, white dwarfs evolve within the galactic environment, where the distribution of surrounding dark matter depends on the mass profile of the galaxy. For G117-B15A, its galactic position is analogous to the Sun’s location within the Milky Way. Hence, we approximate its local dark matter density using the Solar System’s value:$\rho_\chi = 0.3$ $ \text{GeV}/\text{cm}^3$ . Assuming that interstellar diffuse dark matter follows a Maxwell-Boltzmann velocity distribution, we can express its distribution in the rest frame of the white dwarf as follows\cite{busoni2017evaporation,bell2021improved}:
\begin{widetext}
\begin{equation}
    f_{v_*}(u_\chi) du_\chi=\frac{u_\chi}{v_*}\sqrt{\frac{3}{2\pi v_d^2}}\left(\exp \left[-\frac{3(u_\chi-v_*)^2}{2v_d^2} \right]-\exp \left[-\frac{3(u_\chi+v_*)^2}{2v_d^2} \right]  \right) du_\chi ,
\end{equation}
\end{widetext}
where $u_\chi$ represents the velocity of diffuse dark matter far from the white dwarf, $v_\text{d}$ is the dispersion velocity of the galactic dark halo, and $v_*$ denotes the velocity of the white dwarf in the galaxy's rest frame. Assuming the environment of G117-B15A is similar to that of the Solar System, we adopt the values: $v_\text{d} = 270$ km/s and $v_* = 220$ km/s. 

Diffuse dark matter is gravitationally attracted by the white dwarf, accumulating within its interior and interacting with electrons via elastic collision. Under the unsaturated capture condition, the gravitationally accumulated dark matter density is given by\cite{press1985capture,gould1987resonant}:
\begin{equation}
     n_\chi(w_\chi,r)dw_\chi =  \frac{ \sqrt{v_{\text{esc}}(r)^2+u_\chi^2} f_{v_*}(u_\chi)du_\chi}{u_\chi} ,
\end{equation}
where $v_{\text{esc}}(r)$ is the escape velocity at a radial distance  $r$  from the center of the white dwarf. 

In the white dwarf rest frame (the "WD" frame), the scattering condition for dark matter after a collision is conveniently expressed as a velocity exceeding the escape velocity,  $\mathscr{C}_{\text{sca}}(v)=v_\chi-v_{\text{esc}}$. In contrast, expressing this condition in the center-of-momentum (CM) frame of the collision system is less straightforward.   Instead, we use the condition that the outgoing energy of dark matter must be less than the gravitational potential energy: 
\begin{equation}
    \mathscr{C}_{\text{sca}}(\cos\varphi^{\text{CM}},\varphi^{\text{CM}}) =E_\chi'^{\text{WD}}-\frac{1}{\sqrt{1-v_{\text{esc}}^2(r)}}m_\chi,
\end{equation}
where $(\cos\varphi^{\text{CM}},\varphi^{\text{CM}})$ represent the scattering angles of the outgoing dark matter,  $E_\chi'^{\text{WD}} $ denotes the outgoing energy of dark matter, which depends on its scattering angles.  The method for solving $E_\chi'^{\text{WD}} $ is provided in Appendix  \ref{approx_A}. Note that the superscript "CM" and "WD" indicates quantities defined in the center-of-momentum frame and the white dwarf rest frame, respectively. 
The cross-section of dark matter-electron interactions under these conditions can be expressed as:
\begin{widetext}
    \begin{equation}
    \left. \int dv_\chi \frac{d\sigma}{dv_\chi}\right|^{\text{WD}} \Theta(\mathscr{C}_{\text{sca}}(v)) = \int d\cos\theta^{\text{CM}}d\varphi^{\text{CM}} \frac{d\sigma}{d\cos\theta^{\text{CM}} d\varphi^{\text{CM}}} \Theta(\mathscr{C}_{\text{sca}}(\cos\varphi^{\text{CM}},\varphi^{\text{CM}})),
    \end{equation}
\end{widetext}
where $v_\chi$ is the velocity of dark matter after collision. An isotropic differential cross-section consistent with direct detection experiments is adopted: 
\begin{equation}
    \frac{d\sigma}{d\cos\theta^{\text{CM}} d\varphi^{\text{CM}}} = \frac{\sigma_0}{4\pi}|F_{\text{DM}}(q)|^2,
\end{equation}
where $\sigma_0$ is the cross-section for momentum transfer at the atomic scale, and $q$ is the momentum transfer during the collision. In this study, the form factor $|F_{\text{DM}}(q)|^2$ is assumed to be $1$ for heavy mediators.
Due to the high degeneracy of electrons inside white dwarfs, the Pauli exclusion principle suppresses the generation of final electron states. This suppression is represented by the factor $1-f_{\text{FD}}(E_\chi'^{\text{WD}},r)$. 
Combining these discussions, the scattering rate of dark matter in the white dwarf is given by :
\begin{widetext}
    \begin{equation}
            \begin{aligned}
     S^{\text{ngeo}} & =  \int 4\pi r^2 dr \int du_\text{e} dw_\chi F_{\text{møller}} \int d\cos\theta^{\text{CM}}d\varphi^{\text{CM}} \frac{\sigma_0}{4\pi} \Theta(\mathscr{C}_{\text{sca}}(\cos\varphi^{\text{CM}},\varphi^{\text{CM}})) \left(1-f_{\text{FD}}(E_\chi'^{\text{WD}},r)\right)\\
     &= \int_0^{R_*} 4\pi r^2 dr \int \frac{E_\text{e}dE_\text{e} ds}{4\pi^2 \sqrt{E_\chi^2-m_\chi^2}}  f_{\text{FD}}(E_\text{e},r) \int_0^{\infty} \frac{ \sqrt{v_{\text{esc}}(r)^2+u_\chi^2} f_{v_*}(u_\chi)du_\chi}{u_\chi} \\ &\times \int d\cos\theta^{\text{CM}}d\varphi^{\text{CM}} \frac{\sigma_0}{4\pi} \Theta(\mathscr{C}_{\text{sca}}(\cos\varphi^{\text{CM}},\varphi^{\text{CM}})) \frac{\sqrt{(p_\chi\cdot p_\text{e})^2-m_\chi^2 m_\text{e}^2}}{E_\chi E_\text{e}}\left(1-f_{\text{FD}}(E_\chi'^{\text{WD}},r) \right).
        \end{aligned}
    \end{equation}
\end{widetext}
The capture rate is derived by modifying the integration limits in the scattering rate:
\begin{equation}
\begin{aligned}
     C^{\text{ngeo}}  &=  \int 4\pi r^2 dr \int du_\text{e} dw_\chi F_{\text{møller}} \int d\Omega^{\text{CM}} \\
     &\times \frac{\sigma_0}{4\pi} \Theta(\mathscr{C}_{\text{cap}}) \left(1-f_{\text{FD}}(E_\chi'^{\text{WD}},r)\right).
\end{aligned}
\label{eq_C_ngeo}
\end{equation}
Here, for the sake of brevity, the differential solid angle is given by  $d\Omega^{\text{CM}} = d\cos\theta^{\text{CM}}d\varphi^{\text{CM}},$
and the modified boundary is   
\begin{equation*}
    \mathscr{C}_{\text{cap}} = 1/\sqrt{1-v_{\text{esc}}^2(r)}m_\chi- E_\chi'^{\text{WD}}.
\end{equation*}

However, these formulas are valid only under unsaturated capture conditions. When the dark matter-electron scattering cross-section is large, repeated scattering events may lead to overestimations. In such cases, the capture rate is constrained by the geometric limit, which accounts for all dark matter particles gravitationally bound to the white dwarf. The geometric limit is given by\cite{garani2017dark,busoni2017evaporation,bell2021improved}: 
\begin{widetext}
\begin{equation}
    C^{\text{geo}}=\frac{\pi R_*^2}{3v_*} \frac{\rho_\chi}{m_\chi}\left[ ( 3 v_{\text{esc}}^2 (R_*)+ 3 v_*^2+ v_\text{d}^2)\cdot \text{Erf} \left(\sqrt{\frac{3}{2}}\frac{v_*}{v_\text{d}} \right) +\sqrt{\frac{6}{\pi}} v_*v_d\cdot \exp \left( -\frac{3v_*^2}{2v_\text{d}^2} \right) \right] .
\end{equation}
\end{widetext}
Between the unsaturated and geometric limits, dark matter within the white dwarf undergoes scattering attenuation and multiple scattering, leading to highly nonlinear behavior. The capture rate in this regime can be simplified as: 
\begin{equation}
    C_* = min\{C^{\text{geo}},C^{\text{ngeo}}\}.
\end{equation}
When the capture rate reaches the geometric limit, it becomes constant. In contrast, the scattering rate decreases as it approaches the geometric limit. This reduction occurs because sufficiently large cross-sections enable multiple scatterings, fully capturing dark matter and preventing it from escaping the white dwarf. The condition for the attenuation process arises when dark matter can no longer traverse the star without undergoing collisions, described by: 
\begin{equation}
    C^{\text{geo}} = C^{\text{ngeo}}+S^{\text{ngeo}}.
    \label{geo_nor_condition}
\end{equation}
Accordingly, the scattering rate can be calculated separately in two regions:
\begin{equation}
    S_* = \left\{ \begin{aligned} &S^{\text{ngeo}},\quad & C^{\text{ngeo}}+S^{\text{ngeo}} \leq C^{\text{geo}}\\ &C^{\text{geo}}-C^{\text{ngeo}},\quad & C^{\text{ngeo}}+S^{\text{ngeo}} > C^{\text{geo}} \end{aligned} \right. .
\end{equation}
\begin{figure}
    \centering
    \includegraphics[width=1.0\linewidth]{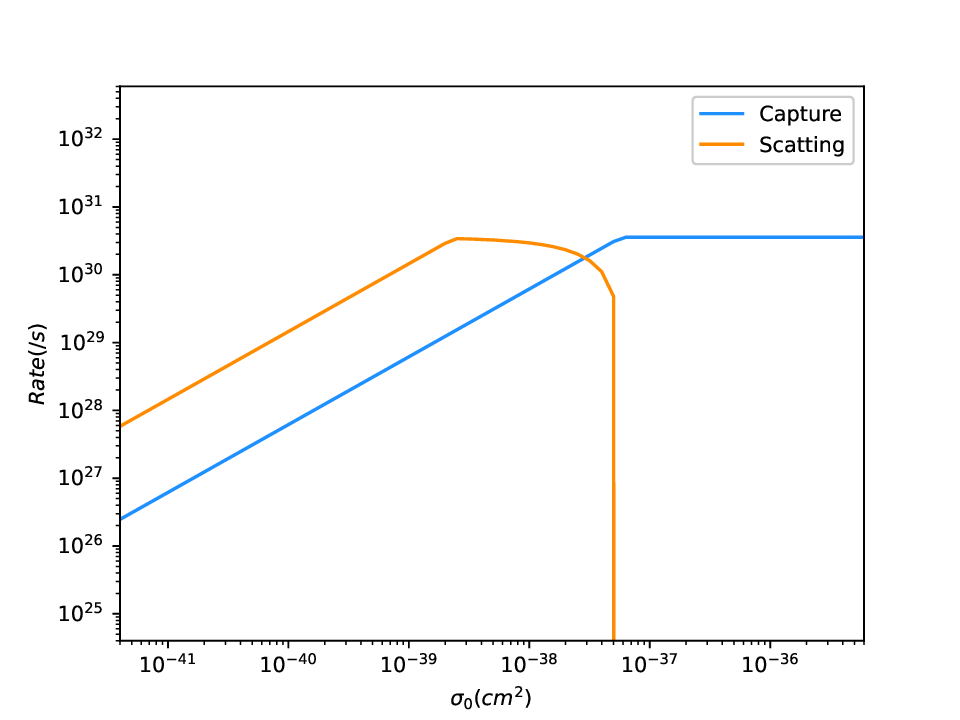}
    \caption{The scattering rate and capture rate depend on the dark matter-electron collision cross-section. The blue solid line corresponds to the capture rate, while the yellow solid line represents the scattering rate. The dark matter mass is fixed at 1 MeV. }
    \label{Scatter_cap_fig}
\end{figure}

Fig. \ref{Scatter_cap_fig} shows the scattering and capture rates as functions of the dark matter-electron collision cross-section for a dark matter mass of 1 MeV.  When the cross-section falls below  $10^{-39}$ $\text{cm}^2$, the unsaturated condition is satisfied, and both the capture rate and scattering rate are positively correlated. As the cross-section approaches $10^{-39}$ $\text{cm}^2$, multiple collisions occur as dark matter passes through the white dwarf. In this regime, dark matter scattered during its first collision may undergo subsequent collisions that lead to its capture, causing the scattering rate to decrease. 
When the cross-section exceeds $3\times 10^{-38}$ $\text{cm}^2$, dark matter undergoes sufficient collisions within the white dwarf to be fully captured, reaching the geometric capture condition. Beyond this point, the capture rate remains constant, constrained by the upper limit set by the gravitational attraction of interstellar dark matter to the white dwarf. Conversely, the scattering rate significantly diminishes. 
However, in cases where the cross-section is sufficiently large to allow multiple collisions within the white dwarf, the actual process becomes highly nonlinear. Our simplified model skips the complex of the region. In the following discussion, we focus on the scenario where dark matter undergoes only a single collision within the white dwarf, as defined by the condition \textit{i.e.} equation (\ref{geo_nor_condition}). 

\subsection{Evaporation and annihilation processes}

The evaporation process involves dark matter \textbf{in the white dwarf's halo} colliding with electrons in the white dwarf interior, gaining sufficient energy to escape the gravitational potential well. Like the scattering process, evaporation considers only those collisions where the post-collision velocity of dark matter exceeds the escape velocity, $\mathscr{C}_{\text{eva}}(v)=v_\chi-v_{\text{esc}}$. Under the unsaturated condition, the evaporation rate can be expressed as follows:

\begin{equation}
    \begin{aligned}
         E^{\text{ngeo}}  &=  \int 4\pi r^2 dr \int du_\text{e} dw_\chi F_{\text{møller}} \int d\Omega^{\text{CM}} \\ &\times \frac{\sigma_0}{4\pi} \Theta(\mathscr{C}_{\text{eva}}) \left(1-f_{\text{FD}}(E_\chi'^{\text{WD}},r)\right).
    \end{aligned}
\end{equation}

The Møller flux $F_{\text{møller}}$ includes the electron distribution within the white dwarf $n_\text{e}(u_\text{e},r)$, as detailed in Section \ref{section_WDstructure}, and the dark matter distribution \textbf{in the white dwarf's halo} $n_\chi^{\text{Halo}}(w_\chi,r)$. Assuming a Maxwell-Boltzmann distribution for \textbf{captured dark matter particles} \cite{gould1987weakly,garani2019new,garani2017dark}, the explicit form of $n_\chi^{\text{Halo}}(w_\chi,r)$ is given by:
\begin{equation}
    n_\chi^{\text{Halo}}(w,r) = N_\chi f_\chi^G(r) f_\chi(w,r),
\end{equation}
where $N_\chi$ denotes the total number of dark matter particles captured by the white dwarf, determined by the balance between capture, evaporation and annihilation processes discussed in next subsection \ref{section_equi}. The term $f_\chi(w,r)$ represents \textbf{the truncated thermal equilibrium distribution}, which accounts for the escape velocity, and $f_\chi^G(r)$ describes the radial distribution influenced by the white dwarf's gravitational potential.
\textbf{The truncated thermal equilibrium distribution} is given by\cite{gould1987weakly,garani2019new,garani2017dark}:
\begin{widetext}
    \begin{equation}
    f_\chi(w,r)=\frac{1}{\pi^{3/2}}\left( \frac{m_\chi}{2kT_*} \right)^{3/2}\frac{\exp(-\frac{m_\chi w^2}{2kT_*})\Theta(v_{\text{esc}}(r)-w)}{\text{Erf}\left(\sqrt{\frac{m_\chi v_{\text{esc}}^2(r)}{2kT_*}} \right)-\frac{2}{\sqrt{\pi}}\sqrt{\frac{m_\chi v_{\text{esc}}^2(r)}{2kT_*}}\exp \left( -\frac{m_\chi v_{\text{esc}}^2(r)}{2kT_*} \right)}.
    \end{equation}
\end{widetext}

Here,  $\Theta(v_{\text{esc}}(r)-w)$ is the Heaviside step function imposing a velocity cutoff based on the escape velocity. The radial distribution of dark matter within the white dwarf, shaped by gravitational effects, is expressed as: 
\begin{equation}
    f_\chi^G(r) =  \frac{1}{r_\chi^3\sqrt{\pi}^3}\exp \left(-\frac{r^2}{r_\chi^2}\right)
\end{equation}
where $r_\chi= \sqrt{3kT_*/2\pi G\rho_*m_\chi}$, the thermalization radius of dark matter, characterizes the spatial extent of its distribution under the influence of the white dwarf's gravitational field.

\textbf{Notably, while the captured dark matter halo distribution provides a reasonable approximation in the regime of heavy dark matter particles, our focus lies within the transitional mass range between heavy and light dark matter species, specifically ($10^{-3}$ MeV, $8$ MeV). As quantified in Appendix \ref{approx_B}, this approximation may lead to an overestimation of results while remaining within the same order of magnitude. However, such approximation completely breaks down when extended to the sub-keV regime($<10^{-3}$ MeV).}

 The dark matter annihilation process occurs \textbf{in the white dwarf's halo.} Hence, The annihilation rate is expressed as:
\begin{equation}
    A =  \int 4\pi r^2 dr \int du_\chi dw_\chi    n_\chi^{\text{Halo}}(u,r) n_\chi^{\text{Halo}}(w,r)\langle \sigma_{\chi \chi} v\rangle.
\end{equation}
For the dark matter annihilation cross-section, we adopt the widely used empirical value: $\langle \sigma_{\chi \chi} v \rangle \approx 3\times 10^{-26}$ $ \text{cm}^3/\text{s}$.

\subsection{Equilibrium of capture, evaporation and annihilation}
\label{section_equi}

The capture process leads to the accumulation of dark matter inside the white dwarf, while evaporation and annihilation contribute to its depletion. The total number of \textbf{captured dark matter particles}, $N_\chi$, evolves according to the following equation:
\begin{equation}
\begin{aligned}
    \frac{dN_\chi(t)}{dt} &= C-E(N_\chi(t))-A( N_\chi(t)^2)\\
    & = C-\hat{E} N_\chi(t)-\hat{A} N_\chi(t)^2.
\end{aligned}  
\end{equation}
Here, $C$,$ E$ and $A$ represent the rates of capture, evaporation, and annihilation, respectively, all of which depend on $N_\chi$. $\hat{E}$ and $\hat{A}$ are the coefficients obtained by factoring out $N_\chi$, defined as:$\hat{E} = E/N_\chi,\hat{A} = A/N_\chi^2$  .
Assuming that these rates remain constant over the white dwarf’s lifetime, the accumulated dark matter at time t can be solved as\cite{gaisser1986limits,griest1987cosmic}:
\begin{equation}
    N_\chi(t) = \frac{C \tanh \left( t /t_{\text{eq}} \right)}{1/t_{\text{eq}}  +\hat{E}/2 \tanh \left( t /t_{\text{eq}} \right)}.
\end{equation}
From this equation, it follows that the accumulated dark matter reaches a steady value after the equilibrium time: 
\begin{equation}
   t_{\text{eq}}= 1/\sqrt{\hat{A}*C+\hat{E}^2/4}.
\end{equation}
\section{Energy flux of Scattering, Capture, Evaporation and Annihilation}
\label{section:4 enegry flux}

\subsection{Energy flux of  capture and scattering}
The process of capture represents a net energy input, as dark matter must lose kinetic energy through collisions to be captured.  After capture, dark matter continues to collide with the stellar material, fully thermalizing and forming a dark halo. The energy input into the white dwarf due to capture, $E_{\text{in}}^{\text{cap}}$, can be expressed as the difference between the pre-capture energy, $E_{\text{in}}$, and the the post-thermalization energy, $E_{\text{eq}}$:
\begin{equation}
    E^{\text{cap}}_{\text{in}} = E_{\text{in}}-E_{\text{eq}}.
\end{equation}
Under the unsaturated capture condition, the pre-capture energy of dark matter is:
\begin{equation}
    E_{\text{in}}^{\text{ngeo}} = \int du_\chi \frac{dC^{\text{ngeo}}}{ du_\chi}(\frac{1}{2}m_\chi u_\chi^2),
\end{equation}
where $C^{\text{ngeo}}$ is defined by equation  (\ref{eq_C_ngeo}). The energy of the thermalized \textbf{white dwarf's halo} is given by :
\begin{equation}
\begin{aligned}
      E_{\text{eq}} &= C_*\cdot \int 4\pi r^2dr 4\pi w^2dw \\
      &\times \left(\frac{1}{2}m_\chi w^2+V_\text{G}(r)\right) n_\chi^{\text{Halo}}(w,r),
\end{aligned}
\end{equation}
where $V_\text{G}(r) = -1/2m_\chi v_{\text{esc}}^2$  is the  gravitational potential of  \textbf{captured dark matter particles}. 
The energy change during the scattering process is challenging to quantify, as it can either increase or decrease the white dwarf’s energy. However, in the cases considered here, scattering typically leads to a net energy outflow. Under the unsaturated condition, this energy outflow is expressed as: 

\begin{equation}
\begin{aligned}
    (E_{\text{out}}^{\text{sca}})^{\text{ngeo}} &= \int 4\pi r^2 dr \int du_e dw_\chi F_{\text{møller}} \int d\Omega^{\text{CM}} \\ 
    & \times \frac{\sigma_0}{4\pi} \Theta(\mathscr{C}_{\text{sca}}) \left(1-f_{\text{FD}}(E_\chi'^{\text{WD}},r)\right)\Delta E_\chi^{\text{sca}},
\end{aligned}
\end{equation}
where $\Delta E_\chi^{\text{sca}} = E_\chi'^{\text{WD}}-E_\chi^{\text{WD}}$ represents the energy gained by dark matter during the scattering process, and $E_\chi^{\text{WD}}$ is the pre-collision energy of dark matter.
As the cross-section increases, dark matter undergoes multiple scatterings.  In this regime, the expressions for $E_{\text{in}}^{\text{ngeo}}$ and $(E_{\text{out}}^{\text{sca}})^{\text{ngeo}}$ under unsaturated conditions are no longer applicable. Specifically, reaching the geometric limit, the capture energy $E_{\text{in}}^{\text{ngeo}}$ becomes constant, while the scattering energy outflow vanishes ($E_{\text{out}}^{\text{sca}} = 0$). In the intermediate regime between unsaturated and geometric conditions, we approximate the energy transfer using the average energy per collision: 
\begin{equation}
    E^{\text{cap}}_{\text{in}} = C_* \frac{(E^{\text{cap}}_{\text{in}})^{\text{ngeo}}}{C^{\text{ngeo}}}, \quad E^{\text{sca}}_{\text{out}} = S_* \frac{(E^{\text{sca}}_{\text{out}})^{\text{ngeo}}}{S^{\text{ngeo}}}.
\end{equation}
\subsection{Energy flux of evaporation and annihilation}

The energy carried away by evaporation for white dwarfs.  Under unsaturated conditions, the energy loss through evaporation can be expressed as:
\begin{equation}
\begin{aligned}
    E_{\text{out}}^{\text{eva}}  &= \int 4\pi r^2 dr \int du_\text{e} dw_\chi F_{\text{møller}} \int d\Omega^{\text{CM}} \\
    &\times \frac{\sigma_0}{4\pi} \Theta(\mathscr{C}_{\text{eva}}) \left(1-f_{\text{FD}}(E_\chi'^{\text{WD}},r)\right)\Delta E_\chi^{\text{eva}},
\end{aligned}
\label{eq_eva}
\end{equation}
where $\Delta E_\chi^{\text{eva}}$ represents the energy transferred during evaporation.
When the cross-section becomes sufficiently large to cause multiple scatterings within the white dwarf, the opacity effect attenuates energy escape. In extreme cases, where the cross-section is exceptionally high, dark matter evaporation behaves similarly to photon radiation, with energy escaping only from the outermost layers of the star. 

\textbf{The annihilation process releases energy equivalent to the total energy of two particles.} The energy released can be given by: 
\begin{equation}
    E^{\text{anni}}_{\text{in}} =   \int  dw du  \frac{dA}{dudw}(\gamma(w)+\gamma(u))m_\chi,
\end{equation}
where $\gamma(u)=1/\sqrt{1-u^2}$ is the Lorentz factor for the velocity $u$.

\subsection{Total energy flux}

Taking into account the energy contributions from scattering, capture, evaporation, and annihilation within the white dwarf, their net effect on the star’s cooling can be expressed as: 
\begin{equation}
    L_\chi = E^{\text{sca}}_{\text{out}}-E^{\text{cap}}_{\text{in}}+E^{\text{eva}}_{\text{out}}-E^{\text{anni}}_{\text{in}}.
\end{equation}
\begin{figure}
    \centering
    \includegraphics[width=1.0\linewidth]{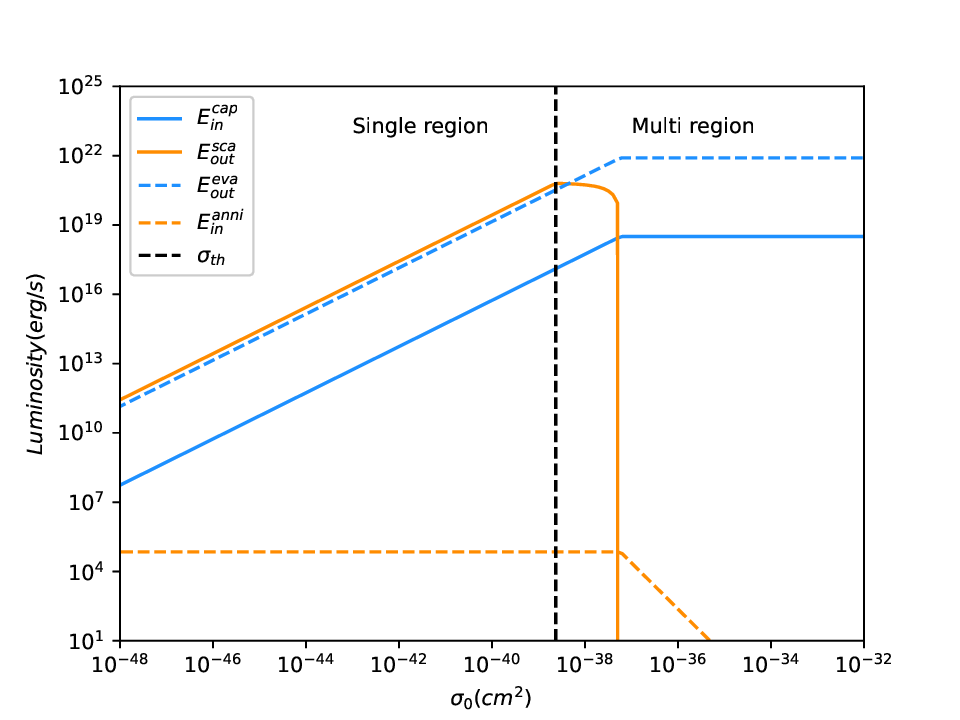}
    \caption{The scattering energy, capture energy, evaporation energy, and annihilation energy as function of  the dark matter-electron collision cross section for a dark matter mass of  $ 1$ $\text{MeV}/\text{c}^2$.  }
    \label{fig_energy_flux}
\end{figure}

Fig. \ref{fig_energy_flux} shows how the energy transfer during scattering, capture, evaporation, and annihilation processes depends on the scattering cross-section with $m_\chi =1 $ $\text{MeV}/\text{c}^2$. The black dashed line marks the single-collision threshold $\sigma_{\text{th}}$. For cross-sections below $\sigma_{\text{th}}$, dark matter undergoes single collisions within the white dwarf. When the cross-section exceeds $\sigma_{\text{th}}$, multiple collisions become significant. 

The yellow solid line represents the energy output from the scattering process, which decreases near the threshold $\sigma_{\text{th}}$. This decrease occurs because dark matter that escapes after a single collision may undergo additional collisions, leading to eventual capture. This behavior aligns with the trend observed in the scattering rate. The blue solid line corresponds to the energy input from the capture process, which reaches its maximum value under the geometric capture condition, consistent with the behavior of the capture rate. 

The green solid line depicts the energy output from the evaporation process. For cross-sections $\sigma_0<\sigma_{\text{th}}$, the energy carried away by evaporation is comparable to that lost through the scattering process. However, it is important to note that the evaporation energy equation (\ref{eq_eva}) does not fully apply in regions where multiple collisions occur. In these cases, \textbf{captured dark matter particles} lose kinetic energy due to repeated collisions while attempting to gain escape velocity. When accounting for internal evaporation, opacity effects must be included, reducing the actual energy carried away compared to what is shown in Fig. \ref{fig_energy_flux}. As this study does not focus on these regions, the conclusions remain unaffected. 

The yellow dashed line represents the energy input into the white dwarf from dark matter annihilation. For small dark matter masses (1 MeV), the annihilation energy input is negligible. However, as the dark matter mass increases, the energy released from annihilation becomes a dominant contributor to the total energy flow.

\section{Result and Discussion}
\label{section:5 result}

When the cross-section exceeds the single-collision threshold $\sigma_{\text{th}}$, multiple collision effects do not increase the energy carried away by dark matter from the white dwarf interior. However, the highly nonlinear nature of this process makes it challenging to model theoretically. Consequently, our calculations are restricted to the region where the unsaturated condition holds.

\begin{figure}
    \centering
    \includegraphics[width=1.0\linewidth]{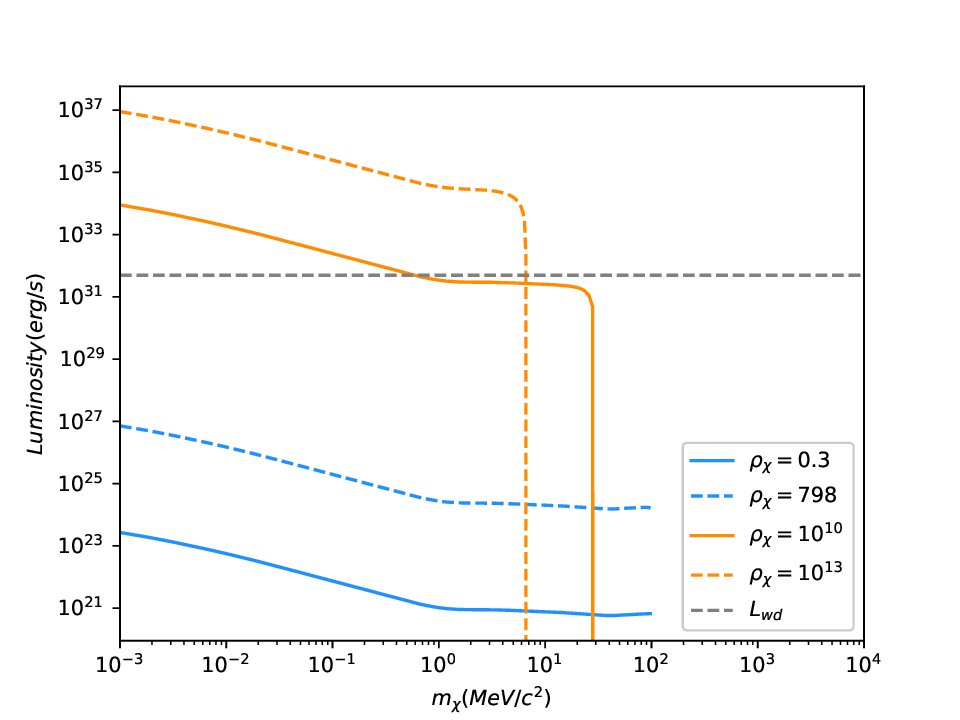}
    \caption{The maximum cooling luminosity of a white dwarf induce by interstellar dark matter via scattering, capture, evaporation, and annihilation processes, as a function of the dark matter mass. The blue solid line represents the maximum cooling luminosity derived from observational data of G117-B15A at an interstellar dark matter density of  $\rho_\chi = 0.3$ $\text{GeV}/\text{cm}^3$. The blue dashed line, yellow solid line, and yellow dashed line correspond to the maximum cooling luminosity when the interstellar dark matter density around G117-B15A is increased to $\rho_\chi=798$ $\text{GeV}/\text{cm}^3$, $10^{10}$ $\text{GeV}/\text{cm}^3$and $10^{13}$ $\text{GeV}/\text{cm}^3$, respectively. The gray dashed line indicates the cooling luminosity threshold ($L_{th}=4.75L_{wd}$), which represents the observational constraint on dark matter-induced cooling luminosity for G117-B15A. }
    \label{fig_L_max}
\end{figure}

In the unsaturated region, the energy carried away by dark matter is positively correlated with the dark matter-electron interaction cross-section, as shown in Fig. \ref{fig_energy_flux}. Using observational data from the pulsating white dwarf G117-B15A, we plot the relationship between the maximum energy carried away by dark matter and the dark matter mass in Fig. \ref{fig_L_max}. 

In Fig. \ref{fig_L_max}, the blue solid line represents the maximum cooling luminosity caused by dark matter at the local solar dark matter density of $\rho_\chi = 0.3$ $\text{GeV}/\text{cm}^3$. The gray solid line indicates the cooling luminosity threshold $4.75L_{wd}$, derived from the observational data of G117-B15A, which serves as an upper limit on dark matter-induced cooling luminosity.  If the calculated maximum cooling luminosity is below this threshold, the sensitivity of the proposed model is sufficient to constrain dark matter. 

Clearly, under solar system-like conditions, G117-B15A cannot provide meaningful constraints. Currently, two potential methods could improve these results. The first involves improving observational data and developing more refined models of pulsating white dwarfs to lower the dark matter cooling luminosity threshold. Lowering the threshold is reflected as a downward shift of the gray dashed line $L_{\text{th}}$. The second method is to identify sources of pulsating white dwarfs in regions with higher dark matter densities to enhance the cooling luminosity caused by dark matter. 

\begin{figure}
    \centering
    \includegraphics[width=1.0\linewidth]{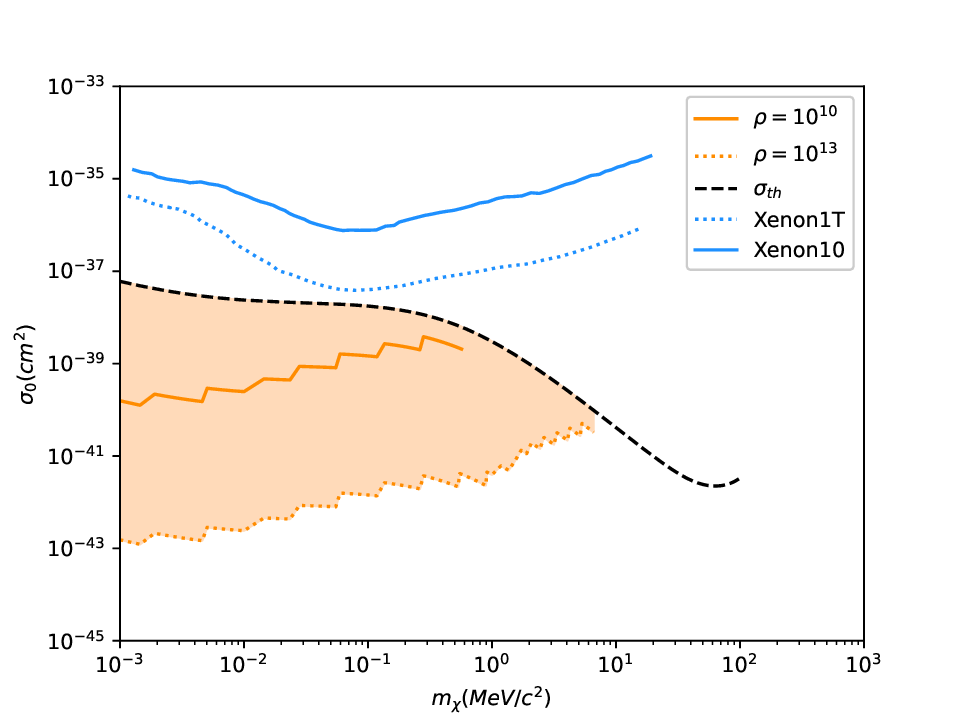}
    \caption{The constraint regions on the dark matter-electron cross-section derived from the observational data of G117-B15A under Galactic Center dark matter densities. The orange solid and dashed lines correspond to constraints at dark matter densities of $\rho_\chi= 10^{10}$ $\text{GeV}/\text{cm}^3$ and $10^{13}$ $\text{GeV}/\text{cm}^3$, respectively. The black dashed line represents the theoretical threshold for dark matter constraints derived from G117-B15A observational data. Constraints proposed in this study must surpass this threshold to be considered valid. The blue solid and dashed lines indicate the cross-section constraints obtained by the Xenon10 and Xenon1T experiments, respectively, based on solar reflection \cite{emken2022solar}. }
    \label{fig_constraint}
\end{figure}

We attempt to calculate the cooling luminosity of white dwarfs in several candidate regions with enriched dark matter densities.  The blue dashed line corresponds to the maximum cooling luminosity derived under the core dark matter density of $\rho_\chi = 798$ $\text{GeV}/\text{cm}^3$ in the M4 globular cluster core (assuming a contracted NFW profile \cite{mccullough2010capture}). Such a density is insufficient to provide constraints with the current quality of observational data, and future improvements in pulsating white dwarf theory and observations may be required to reduce the threshold. The yellow solid and dashed lines correspond to the Galactic Center dark matter densities (assuming an NFW profile \cite{navarro1996structure}), with $\rho_\chi= 10^{10}$ $\text{GeV}/\text{cm}^3$ and $10^{13}$ $\text{GeV}/\text{cm}^3$, respectively. In these cases, the dark matter-induced cooling luminosity exceeds the threshold, enabling constraints, as shown in Fig. \ref{fig_constraint}.

In Fig. \ref{fig_constraint}, the yellow solid and dashed lines depict constraints on the dark matter-electron scattering cross-section at Galactic Center dark matter densities of $\rho_\chi= 10^{10}$ $\text{GeV}/\text{cm}^3$ and $10^{13}$ $\text{GeV}/\text{cm}^3$, respectively, derived using different NFW profiles. These constraints exceed those obtained by Xenon10 and Xenon1T experiments via solar reflection of dark matter \cite{an2018directly,an2021solar,emken2022solar}.

While obtaining observational data of pulsating white dwarfs in the Galactic Center remains challenging, future advancements in instrumentation and observational techniques may make such measurements feasible. The extremely high dark matter densities in the Galactic Center result in cooling luminosities comparable to the photon luminosities of white dwarfs. This leads to significant effects on white dwarf evolution, potentially altering their thermal history and observable properties. 

\section{Conclusion}
\label{section:6 conclusion}

Theoretical and observational studies of pulsating white dwarf modes provide constraints on dark matter-induced cooling processes in white dwarfs. Motivated by this, we numerically computed the cooling luminosity caused by interstellar sub-MeV dark matter. Our analysis considered scattering, capture, evaporation, annihilation, and the equilibrium processes involving white dwarfs and surrounding dark matter, with detailed calculations of the energy input and output from each process.

We incorporated the degeneracy of electrons in the white dwarf and the relativistic effects of their collisions with dark matter. Our results show that the cooling luminosity of G117-B15A caused by interstellar dark matter is insufficient to significantly impact its evolution. Observational data and theoretical models of G117-B15A do not provide meaningful constraints on dark matter under these conditions.

To explore stronger constraints, we investigated regions with high dark matter densities, such as the Galactic Center. Under the NFW profile, our calculations suggest that significant constraints on dark matter properties can be achieved.

Future advancements in observational instruments and techniques may facilitate the identification of pulsating white dwarfs in the Galactic Center, and further development of pulsation mode theories could lower the threshold for dark matter-electron scattering cross-section constraints. On the other hand, our findings reveal that interstellar dark matter has a negligible effect on the cooling of white dwarfs outside the Galactic Center. However, under Galactic Center conditions, dark matter must be considered in evolutionary models of white dwarfs due to its substantial impact on their cooling and thermal history.

\acknowledgments
This work is supported by the National Key R\&D Program of China (Grants No. 2022YFF0503304), the National Natural Science Foundation of China (12373002, 12220101003, 11773075) and the Youth Innovation Promotion Association of Chinese Academy of Sciences (Grant No. 2016288).

\bibliography{refs}

\appendix
\section{Dark matter energy after collision}
\label{approx_A}
\setcounter{equation}{0}
\setcounter{figure}{0}
In the white dwarf rest frame, the 4-momentum of the electron before collision is $p_e^{\text{WD}}=(E_e^{\text{WD}},\vec{p_\text{e}}^{\text{WD}})$, , and that of the dark matter particle is $p_\chi^{\text{WD}}=(E_\chi^{\text{WD}},\vec{p_\chi}^{\text{WD}})$. After the collision, the 4-momentum of the dark matter particle becomes $p_\chi '^{\text{WD}}=(E_\chi '^{\text{WD}},\vec{p}_\chi '^{\text{WD}})$. Based on these initial conditions, the center-of-mass (CM) velocity of the electron-dark matter system, $\vec{v}_{\text{CM}}$, is given by: 
\begin{equation}
    |\vec{v}_{\text{CM}}| = \sqrt{\frac{(\vec{p}_e^{\text{WD}}+\vec{p}_\chi^{\text{WD}})^2}{(E_e^{\text{WD}}+E_\chi^{\text{WD}})^2}} = \sqrt{1-\frac{s}{(E_e^{\text{WD}}+E_\chi^{\text{WD}})}}.
\end{equation}
Here, the superscript “WD” indicates values in the white dwarf rest frame. For simplicity, all derivations in this section are conducted in the CM frame, where the "CM" superscript is omitted. 
In the CM frame, the 4-momentum definitions are reintroduced: the 4-momentum of the electron before the collision is $p_\text{e}=(E_\text{e},\vec{p_\text{e}})$, that of the dark matter particle before collision is $p_\chi=(E_\chi,\vec{p_\chi})$, and the 4-momentum of the dark matter particle after collision is $p_\chi '=(E_\chi ',\vec{p}_\chi ')$.

The Fig. \ref{fig_collision_frame} below illustrates the collision in the CM frame. The velocity of the white dwarf rest frame relative to the CM frame is denoted by $\vec{v}_{\text{WD}}$. The angle $\alpha$ represents the inclination between $\vec{v}_{\text{WD}}$ and the incoming dark matter momentum $\vec{p}_\chi$. The angle $\theta$ measures the separation between the incoming dark matter momentum $\vec{p_\chi}$ and the outgoing dark matter momentum $\vec{p_\chi}'$, while $\varphi$ is the azimuthal angle between the planes defined by 
$(\vec{v}_{\text{WD}},\vec{p}_\chi)$ and $(\vec{p}_\chi,\vec{p}'_\chi)$. Together, $(\theta,\varphi)$ describe the angular coordinates of the outgoing dark matter momentum. 
The velocity of the white dwarf rest frame relative to the CM frame is derived as: 
\begin{equation}
    |\vec{v}_{\text{WD}}| = |-\vec{v}_{\text{CM}}| = \sqrt{1-\frac{s}{(E_e^{\text{WD}}+E_\chi^{\text{WD}})}},
\end{equation}
where $s$ is the Mandelstam variable.
\begin{figure}
    \centering
    \includegraphics[width=1.0\linewidth]{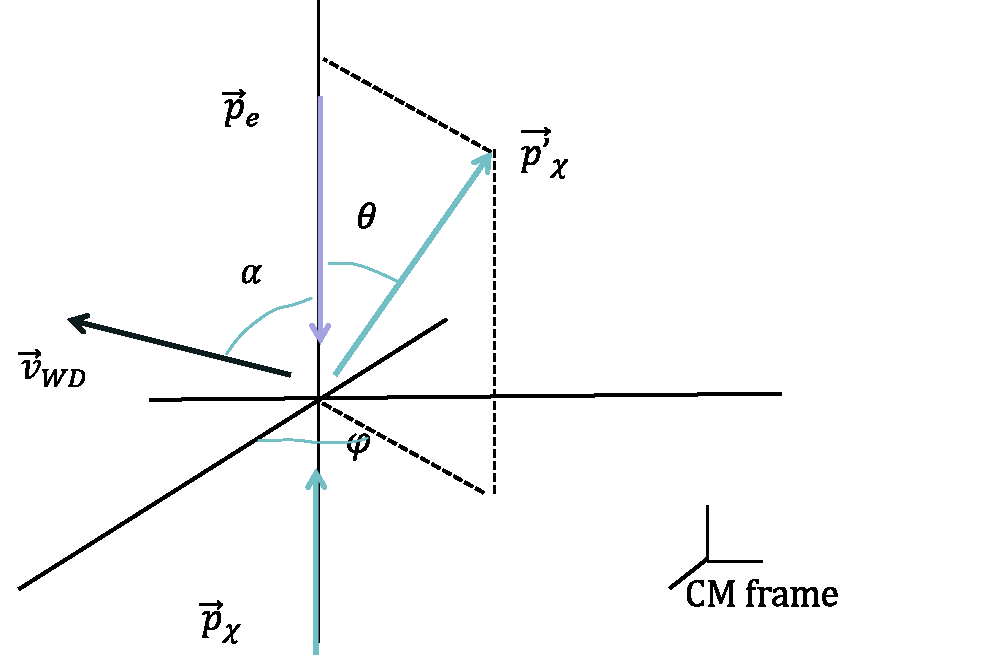}
    \caption{Electron-dark matter collision schematic diagram.}
    \label{fig_collision_frame}
\end{figure}

Our goal is to calculate the energy of the scattered dark matter particle, $E_\chi'^{\text{WD}}$, in the white dwarf rest frame. In the center-of-momentum (CM) frame, the energy of the scattered dark matter, $E_\chi'^{\text{CM}}$, is determined by 4-momentum conservation: 
\begin{equation}
    E_\chi'^{\text{CM}}=\sqrt{m_\chi^2+ \vec{p}_\chi'^2}.
\end{equation}
The corresponding energy in the white dwarf rest frame, $E_\chi'^{\text{WD}}$, can be obtained by performing a Lorentz transformation on the 4-momentum $p_\chi'^{\text{CM}}$. This transformation requires the angle $\gamma$ between the CM frame velocity $\vec{v}_{\text{CM}}$ and the scattered dark matter momentum $\vec{p}_\chi '$, which is given by: 
\begin{equation}
    \cos \gamma = \cos \alpha \cos \theta + \sin \alpha \cdot \sin \theta \cos \varphi .
\end{equation}
The velocity of the dark matter particle in the white dwarf rest frame is then:
\begin{equation}
    \vec{v}_\chi'^{\text{WD}} = L(\vec{v}_{\text{WD}})\vec{v}_\chi'^{\text{CM}}.
\end{equation}

To simplify the calculation of $L(\vec{v}_{\text{WD}})$, we align $\vec{v}_{\text{WD}}$ along the $x$-axis and construct a right-handed coordinate system where the $y$-axis is defined by the normal to the plane formed by $\vec{v}_{\text{WD}}$ and $\vec{p}_\chi '$. In this coordinate system, the  $L(\vec{v}_{\text{WD}})$ is the solution of  the equations:
\begin{align} (v_{\chi}'^{\text{WD}})_x & = \frac{(v_\chi '^{\text{CM}})_x -|\vec{v}_{\text{WD}}|}{1-(v_\chi '^{\text{CM}})_x \cdot |\vec{v}_{\text{WD}}|},\\ (v_{\chi}'^{\text{WD}})_y &= \frac{\sqrt{1-|\vec{v}_{\text{WD}}|^2}}{1-|\vec{v}_{\text{WD}}| \cdot (v_\chi '^{\text{CM}})_x} (v_\chi '^{\text{CM}})_y, \\ (v_{\chi}'^{\text{WD}})_z & = \frac{\sqrt{1-|\vec{v}_{\text{WD}}|^2}}{1-|\vec{v}_{\text{WD}}| \cdot (v_\chi '^{\text{CM}})_x} (v_\chi '^{\text{CM}})_z, \\ \vec{v}_\chi '^{\text{CM}} &= \left( |\vec{v}_\chi '^{\text{CM}}| \cos \gamma,\quad 0, \quad |\vec{v}_\chi '^{\text{CM}}|\sin \gamma \right) ,\\ | \vec{v}_\chi '^{\text{CM}}| &= | \vec{v}_\chi ^{\text{C}}| .\end{align} 

In the CM frame, the velocity of the incoming dark matter particle is expressed as: 
\begin{equation}
    |\vec{v}_\chi^{\text{CM}}| = \frac{|\vec{p}_\chi^{\text{CM}}|}{E_\chi^{\text{CM}}},
\end{equation}
where $|\vec{p}_\chi^{\text{CM}}|$ and $E_\chi^{\text{CM}}$ are functions of the Mandelstam variable $s$:
\begin{align} |\vec{p}_\chi^{\text{CM}}| &= \frac{1}{2\sqrt{s}}\sqrt{s^2-2s(m_e^2+m_\chi^2)+(m_e^2-m_\chi^2)^2}, \\ E_\chi^{\text{CM}} &= \frac{1}{2\sqrt{s}}(s+m_\chi^2-m_e^2). \end{align}
The outgoing energy $E_\chi'^{WD}$ corresponds to the 0th component of the outgoing 4-momentum and is given by: 
\begin{equation}
    E_\chi '^{\text{WD}} = \frac{m_\chi}{\sqrt{1-|\vec{v}_\chi '^{\text{WD}}|^2}}.
\end{equation}

This completes the calculation of the outgoing energy. For continuity in the earlier discussion, the methods for calculating the scattering angles $(\alpha,\theta,\varphi)$ in the diagram were not provided but are detailed below. We begin with the angles $(\theta,\varphi)$ associated with the outgoing dark matter momentum. The polar angle $\theta$ can be expressed using the Mandelstam variable $t$:
\begin{align} t = (p_\chi-p_\chi ')^2 &= 2m_\chi^2-2(E_\chi^{\text{CM}})^2+2|\vec{p}_\chi^{\text{CM}}|^2 \cdot \cos \theta \\ \cos \theta  &= \frac{t-2m_\chi^2+2(E_\chi^{\text{CM}})^2}{2|\vec{p}_\chi^{\text{CM}}|^2}.\end{align}

The azimuthal angle  $\varphi$ is an independent variable that must be specified separately, as it is not constrained by the scattering process.
Finally, we address the calculation of the angle $\alpha$, which represents the inclination between the velocity of the white dwarf rest frame, $\vec{v}_{\text{WD}}$ , and the incoming dark matter momentum  $\vec{p}_\chi$.
In the white dwarf rest frame, the velocity of the incoming dark matter particle is $\vec{v}^{\text{WD}}_\chi$, while the velocity of the center-of-momentum (CM) frame is $\vec{v}_{\text{CM}}$. The angle between these two velocities, $\phi^{\text{WD}}$, is obtained by comparing two equivalent expressions for $p_\chi \cdot (p_\chi + p_e )$:
\begin{align} p_\chi \cdot (p_\chi +p_e) &= m_\chi^2+\frac{s-m_\chi^2-m_e^2}{2}, \\ p_\chi \cdot (p_\chi +p_e) &= E_\chi^{\text{WD}}(E_\chi^{\text{WD}}+E_e^{\text{WD}})\\ \notag & -|\vec{p}_\chi^{\text{WD}}|\cdot |\vec{p}_\chi^{\text{WD}}+\vec{p}_e^{\text{WD}}|\cos \phi^{\text{WD}} .\end{align}
This yields: 
\begin{equation}
    \cos \phi^{\text{WD}} = \frac{E_\chi^{\text{WD}}(E_\chi^{\text{WD}}+E_e^{\text{WD}})-\frac{s+m_\chi^2-m_e^2}{2}}{\sqrt{(E_\chi^{\text{WD}})^2-m_\chi^2}\cdot \sqrt{(E_e^{\text{WD}}+E_\chi^{\text{WD}})^2-s}}.
\end{equation}
Given $\phi^{\text{WD}}\in [0,\pi]$, the angle can be precisely determined.

In the CM frame, the velocity of the incoming dark matter, $\vec{v}^{\text{CM}}_\chi$, is obtained via a Lorentz transformation: 
\begin{equation}
    \vec{v}_\chi^{\text{CM}} = L(\vec{v}_{\text{CM}})\vec{v}_\chi^{\text{WD}}.
\end{equation}
To simplify the transformation, we align $\vec{v}_{\text{CM}}$ along the x-axis of the white dwarf rest frame, constructing a right-handed coordinate system where the $y$-axis is perpendicular to the plane formed by $\vec{v}_{\text{CM}}$ and $\vec{v}_\chi^{\text{WD}}$. The Lorentz transformation $L(\vec{v}_{\text{CM}})$ is then expressed as: 
\begin{align} (\vec{v}_\chi^{\text{CM}})_x & = \frac{(\vec{v}_\chi ^{\text{WD}})_x -|\vec{v}_{\text{CM}}|}{1-(\vec{v}_\chi ^{\text{WD}})_x \cdot |\vec{v}_{\text{CM}}|},\\ (\vec{v}_\chi^{\text{CM}})_y &= \frac{\sqrt{1-|\vec{v}_{\text{CM}}|^2}}{1-|\vec{v}_{\text{CM}}| \cdot (\vec{v}_\chi ^{\text{WD}})_x} (\vec{v}_\chi ^{\text{WD}})_y, \\ (\vec{v}_\chi^{\text{CM}})_z & = \frac{\sqrt{1-|\vec{v}_{\text{CM}}|^2}}{1-|\vec{v}_{\text{CM}}| \cdot (\vec{v}_\chi ^{\text{WD}})_x} (\vec{v}_\chi ^{\text{WD}})_z, \\ \vec{v}_\chi ^{\text{WD}} &= \left( |\vec{v}_\chi^{\text{WD}}|\cos \phi^{\text{WD}},\quad 0, \quad |\vec{v}_\chi^{\text{WD}}| \sin \phi^{\text{WD}} \right) .\end{align}

Finally, the angle $\alpha$ between the velocity of the white dwarf rest frame and the incoming dark matter momentum in the CM frame is calculated as the angle between $\vec{v}_\chi^{\text{WD}}$ and the negative x-axis: 
 \begin{equation}
     \tan \alpha = -\frac{(\vec{v}_\chi^{\text{CM}})_y}{(\vec{v}_\chi^{\text{CM}})_x}.
 \end{equation}
 By solving for $\alpha, \theta$ and $\varphi$ the outgoing energy $E_\chi'^{\text{WD}}$ can be determined. 

\section{the validity of the captured heavy dark matter halo distribution approximation}
\label{approx_B}

\textbf{
The evolution of diffuse captured dark matter halos presents significant complexity, with galactic dark matter halos being extensively discussed in existing literature. While established models such as the King profile \cite{king1966structure} and double power-law distribution \cite{lisanti2011dark} have been proposed, these approaches inevitably introduce experimental uncertainties \cite{bose2023impact}. Given the distinct dynamical behaviors between stellar-mass compact objects and galactic systems, a simplified distribution model is proposed. We define three key parameters for model evaluation: normalized overlap coefficient, maximum magnification factor, and minimum magnification factor. Comparative analysis between our simplified model and conventional heavy dark matter distribution frameworks establishes the validity domain of existing heavy dark matter models.
}

\subsection{Simplified Captured Dark Matter Halo Model}

\textbf{
The dark matter distribution in captured dark matter halos is modeled using a modified Maxwell-Boltzmann distribution:
}

\begin{align}
    f_{\text{my}}(v,r) &= \frac{1}{\mathcal{N}}\exp\left(-\frac{1/2 m_\chi v^2+\Phi(r)}{k_B T_*} \right) \nonumber\\
    &\times \Theta\left[\Phi(\infty)-(1/2 m_\chi v^2+\Phi(r))\right],
\end{align}

\textbf{
where $\mathcal{N}$ denotes the normalization constant and $\Phi(r)$ represents the gravitational potential. This formulation unifies the zero-potential reference point at the stellar core. For computational tractability, we model the star as a uniform-density sphere with gravitational potential:
}

\begin{equation}
    \Phi(r) = \left\{ \begin{array}{cc} \frac{GM_*m_\chi}{2}\frac{r^2}{R_*^3} & r<R_* \\ -\frac{GM_*m_\chi}{r}+\frac{3GM_*m_\chi}{2R_*} & r>R_* \end{array} \right.,
\end{equation}

\textbf{
where $r$ indicates the radial distance from the stellar center. Notably, the zero-potential reference is defined at the stellar center rather than at infinity.
}

\subsection{Similarity Metric Analysis}

\textbf{
We introduce the normalized overlap coefficient (NOC) to quantify the similarity between two distributions:
}

\begin{equation}
    NOC = \frac{\int drdv \min(f_{\text{my}},f_{\text{hea}})}{\int drdv \max(f_{\text{my}},f_{\text{hea}})}.
\end{equation}
\textbf{
This dimensionless parameter is bounded within $[0,1]$, where unity indicates identical distributions and higher values correspond to greater similarity.
}

\begin{figure}
    \centering
    \includegraphics[width=1.0\linewidth]{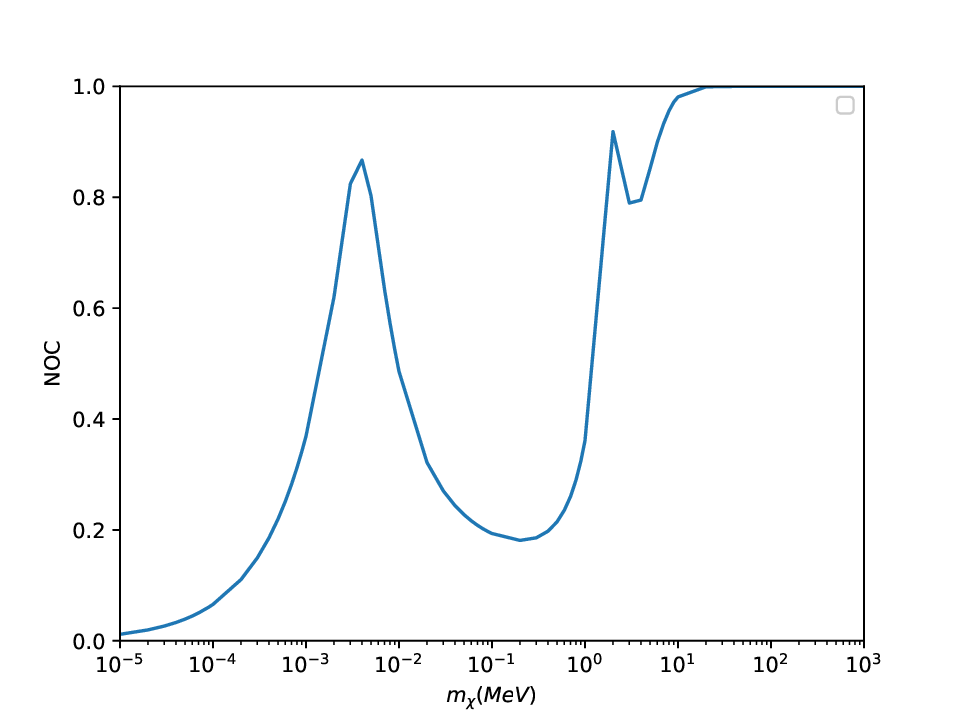}
    \caption{\textbf{The mass-dependent NOC evolution diagram.}}
    \label{fig_NOC_mx_relation}
\end{figure}

\textbf{
Fig. \ref{fig_NOC_mx_relation} demonstrates the mass-dependent NOC evolution between our simplified model and conventional heavy dark matter distributions. Key observations include:
}
\textbf{
\begin{itemize}
    \item Approximation fidelity degrades for particle masses below $8$ MeV
    \item A transient resurgence near $10^{-2}$ MeV
    \item Complete model breakdown below $10^{-3}$ MeV
\end{itemize}
}
\textbf{
To quantify the approximation bounds, we establish inequality constraints:
}

\begin{align}
    \int drdv &f_{\text{my}}(v,r)\mathscr{F}(v,r) \\ &> \min\left[\frac{f_{\text{my}}(v,r)}{f_{\text{hea}}(v,r)}\right]\int drdv f_{\text{hea}}(v,r)\mathscr{F}(v,r), \nonumber \\ 
    \int drdv &f_{\text{my}}(v,r)\mathscr{F}(v,r)  \\&< \max\left[\frac{f_{\text{my}}(v,r)}{f_{\text{hea}}(v,r)}\right]\int drdv f_{\text{hea}}(v,r)\mathscr{F}(v,r) \nonumber,
\end{align}
\textbf{
where $\mathscr{F}$ encapsulates remaining functional components. The extremal magnification factors are defined as:
}
\begin{align}
    \text{Max}\, \text{Mgn} &=  \max\left[\frac{f_{\text{my}}(v,r)}{f_{\text{hea}}(v,r)}\right],\\ \text{Min}\, \text{Mgn} &= \min\left[\frac{f_{\text{my}}(v,r)}{f_{\text{hea}}(v,r)}\right].
\end{align}

\textbf{
These parameters delineate the upper/lower bounds of approximation deviation between conventional heavy dark matter halo distribution models and our simplified framework.
}

\subsection{Comparative Model Evaluation}

\begin{figure}
    \centering
    \includegraphics[width=1.0\linewidth]{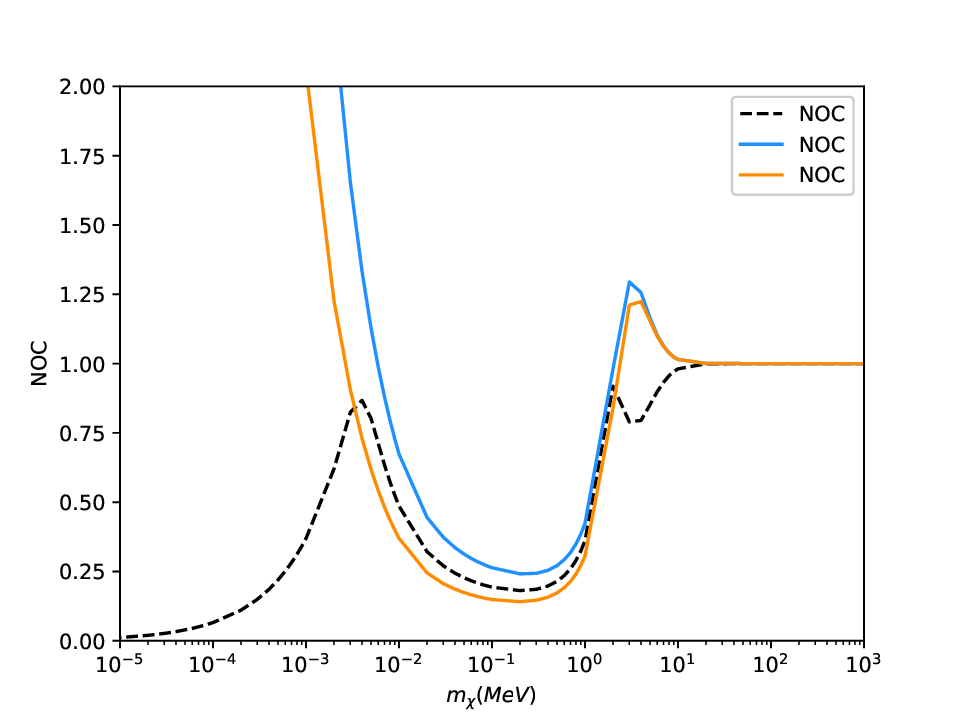}
    \caption{\textbf{The mass-dependent evolution diagram of Max Mgn and Min Mgn.}}
    \label{fig_Max_Min}
\end{figure}

\textbf{
Fig. \ref{fig_Max_Min} reveals the mass-dependent evolution of both Max Mgn and Min Mgn, demonstrating remarkable consistency with the NOC analysis. Our primary focus resides in the transitional mass window ($10^{-3}$ MeV, $8$ MeV), where the simplified model predicts density profiles $0.2-1.0$ times those of conventional captured heavy dark matter halo models. This reveals systematic overestimation inherent to heavy dark matter paradigms, while discrepancies are confined within one order of magnitude. Notably below $3\times 10^{-3}$ MeV the heavy dark matter framework conversely underestimates density distributions, explaining the observed NOC resurgence near $3\times 10^{-3}$ MeV.
}

\subsection{Model Limitations and Extensions}
\textbf{
The simplified framework incorporates non-quadratic potential energy terms neglected in the captured heavy dark matter halo approximations. For improved accuracy, two potential refinements emerge:
}
\textbf{
\begin{itemize}
    \item Spectral decomposition: Expressing the true halo distribution through Boltzmann basis functions with distributed zero-potential references.
    \item Empirical parametrization: Developing phenomenological models calibrated with numerical simulations.
\end{itemize}
}

\textbf{
Despite these possibilities, increased model complexity would compromise computational efficiency. Our comparative analysis confirms the heavy dark matter approximation maintains order-of-magnitude validity within ($10^{-3}$ MeV, $8$ MeV), but suffers complete breakdown below $10^{-3}$ MeV where light dark matter effects dominate.
}

\end{document}